\documentclass[traditabstract]{aa}
\usepackage{graphicx,epsf,textcomp,pstricks,psfrag,amsmath,wasysym,txfonts,longtable,booktabs}
\usepackage{natbib}
\bibpunct{(}{)}{;}{a}{}{,}
\begin{document}

\sloppy


\newcommand{\total}{\operatorname{d}\!}
\renewcommand{\arraystretch}{1.05}
\tabcolsep 3pt

\newcommand{\mum}{\,\hbox{\textmu{}m}}
\newcommand{\mm}{\,\hbox{mm}}
\newcommand{\cm}{\,\hbox{cm}}
\newcommand{\m}{\,\hbox{m}}
\newcommand{\km}{\,\hbox{km}}
\newcommand{\AU}{\,\hbox{AU}}
\newcommand{\pc}{\,\hbox{pc}}
\newcommand{\Kelvin}{\,\hbox{K}}
\newcommand{\second}{\,\hbox{s}}
\newcommand{\yr}{\,\hbox{yr}}
\newcommand{\Myr}{\,\hbox{Myr}}
\newcommand{\Gyr}{\,\hbox{Gyr}}
\newcommand{\rad}{\,\hbox{rad}}
\newcommand{\mJy}{\,\hbox{mJy}}
\newcommand{\magn}{\,\hbox{mag}}
\newcommand{\scmperccm}{\,\hbox{cm}^{2}\hbox{cm}^{-3}}
\newcommand{\persmpers}{\,\hbox{m}^{-2}\hbox{s}^{-1}}
\newcommand{\g}{\,\hbox{g}}
\newcommand{\gperccm}{\,\hbox{gcm}^{-3}}
\newcommand{\ergperg}{\,\hbox{ergg}^{-1}}
\newcommand{\gpers}{\,\hbox{gs}^{-1}}
\newcommand{\kmpers}{\,\hbox{kms}^{-1}}

\title{An improved model of the Edgeworth-Kuiper debris disk}

\bigskip
\author{Christian Vitense
        \and
        Alexander V. Krivov
        \and
        Hiroshi Kobayashi
        \and
        Torsten L\"ohne
       }
\offprints{Ch.~Vitense, \email{vitense@astro.uni-jena.de}}
\institute{Astrophysikalisches Institut, Friedrich-Schiller-Universit\"at Jena,
           Schillerg\"a{\ss}chen~ 2--3, 07745 Jena, Germany
          }
\date{Received 30 November 2011; accepted 07 February 2012}

\abstract
{
In contrast to all other debris disks,
where the dust can be seen via an infrared excess over the stellar photosphere,
the dust emission of the Edgeworth-Kuiper belt (EKB)
eludes remote detection due to the strong foreground emission of the
zodiacal cloud.
In this paper, we access the expected EKB dust disk properties
by modeling.
We treat the debiased population of the known  
transneptunian objects (TNOs)
as parent bodies and generate the dust with our collisional code.
The resulting dust distributions are modified to take into account
the influence of gravitational scattering and resonance trapping by planets
on migrating dust grains as well as the effect of sublimation.
A difficulty with the modeling is that the amount and distribution of dust
are largely determined by sub-kilometer-sized bodies.
These are directly unobservable, and their properties cannot be accessed by
collisional modeling,
because objects larger than $(10\dots60)\m$ in the present-day EKB are not
in a collisional equilibrium.
To place additional constraints,
we use in-situ measurements of the New Horizons spacecraft within $20\AU$.
We show that, to sustain a dust disk consistent with these measurements,  
the TNO population \emph{has} to have a break in the size
distribution at $s\lesssim 70\km$.
However, even this still leaves us with several models that all correctly reproduce a
nearly constant dust impact rates in the region of giant planet orbits
and do not violate the constraints from the non-detection of the EKB  
dust thermal emission by the COBE spacecraft.
The modeled EKB dust disks, which conform to the observational
constraints, can either be transport-dominated or intermediate between
the transport-dominated and collision-dominated regime.
The in-plane optical depth of such disks is $\tau_\parallel(r>10\AU) \sim 10^{-6}$
and their fractional luminosity is $f_d \sim 10^{-7}$.
Planets and sublimation are found to have little effect on dust
impact fluxes and dust thermal emission.
The spectral energy distribution of an EKB analog,
as would be seen from $10\pc$ distance, peaks at wavelengths of
$(40\dots50)\mum$ at $F \approx 0.5\mJy$, which is less than $1\%$ of the
photospheric flux at those wavelengths.
Therefore, EKB analogs cannot be detected with present-day instruments
such as Herschel/PACS.


}

\authorrunning{Vitense et al.}
\titlerunning{An improved model of the Edgeworth-Kuiper debris disk}

\maketitle

\section{Introduction}

The Edgeworth-Kuiper Belt (EKB) with its presumed collisional debris is the main
reservoir of small bodies and dust in the Solar System and constitutes the most prominent
part of the Solar System's debris disk. However, the EKB dust has not been unambiguously
detected so far. The observational evidence for the EKB dust is limited to
scarce in-situ detections of dust in the outer Solar System by a few spacecraft, partly
with uncalibrated ``chance detectors''
\citep{Gurnett-et-al-1997,Landgraf-et-al-2002,Poppe-et-al-2010}.
In addition, there are rough upper limits on the amount of dust from the non-detection
of thermal emission of the EKB dust on a bright zodiacal light foreground
\citep{Backman-et-al-1995}. Given the lack of observational data, one can only
access the properties of the EKB dust by modeling. Such a modeling takes the known
EKB populations to be parent bodies for dust and uses collisional models to
generate dust distributions \citep{Stern-1995,Stern-1996,Vitense-et-al-2010,Kuchner-Stark-2010}.

In our previous paper \citep{Vitense-et-al-2010},
we took the current database of known transneptunian objects (TNOs)
and employed a new algorithm to eliminate the inclination and the distance selection
effects in the known TNO populations and derived expected
parameters of the ``true'' EKB.
Treating the debiased populations
of EKB objects as dust parent bodies, we then produced their dust disk
with our collisional code.

The main goal of this paper is to improve the model by
\citet{Vitense-et-al-2010} in several important respects:
\begin{itemize}
 \item[I.] Although we do not modify the debiasing algorithm and stay with the same ``true'' EKB
 as defined in \cite{Vitense-et-al-2010}, we re-address the question of how the size distribution in the present-day
 EKB that we only know down to sizes of $\sim 10\km$ should be extrapolated down to the dust sizes.
 Accordingly, we present the new collisional code runs that make different assumptions about the amount of objects smaller 
 than $\sim 10\km$ in the current EKB.
 Besides, these new runs include a more realistic material composition (a mixture of ice and 
 astrosilicate in equal fractions)
 and an accurate handling of the cross-section of dust grains.
  This is done in Section~\ref{sec:dust_production}.
 \item[II.] We estimate the influence of planets (resonant trapping and gravitational scattering)
  (Section~\ref{sec:planets}).
 \item[III.] We include the possible effect of ice sublimation
  (Section~\ref{sec:sublimation}).
 \item[IV.] We finally make a detailed comparison of the model with the spacecraft in-situ measurements,
 including the first results of New Horizons \citep{Poppe-et-al-2010,Han-et-al-2011}, as well as with the thermal emission 
 constraints by COBE \citep[][and references therein]{Greaves-Wyatt-2010}.
 This is done in Sections~\ref{sec:in-situ} and~\ref{sec:SED}.
\end{itemize}
Our results are summarized in Section~\ref{sec:conclusion} and discussed in 
Section~\ref{sec:discussion}.

\section{The dust production model}\label{sec:dust_production}

\subsection{The size distribution in the EKB}

We start with general remarks about the size distribution in the EKB and its evolution
since the early phases of the Solar System formation.
Since it is not known how planetesimals in the solar nebula have formed, their primordial
size distribution is unclear. In standard coagulation scenarios,
the bottom-up growth of planetesimals could have resulted in a broad size 
distribution \citep[e.g.,][]{Kenyon-Bromley-2008}, with a more or less constant slope
across all the sizes up to roughly the size of Pluto.
Alternatively, local gravitational instability in turbulent disks would have produced
predominantly big ($\sim 100$~km) planetesimals
\citep{Johansen-et-al-2006,Johansen-et-al-2007,Cuzzi-et-al-2008,Morbidelli-et-al-2009b},
implying a knee in the size distribution at such sizes
which is indicated by several observations
\citep{Bernstein-et-al-2004,Fuentes-Holman-2008,Fraser-Kavelaars-2009,Fuentes-et-al-2009}.
Next, according to the Nice model \citep{Gomes-et-al-2005,Levison-et-al-2008,Morbidelli-2010},
the primordial Kuiper belt was compact (between $15$ and $35\AU$) and massive ($\sim 50$ Earth 
masses). With these parameters, the EKBOs with sizes up to hundreds of kilometers would have
been collisionally processed by the time of the Late Heavy Bombardment (LHB) in
$\approx 800\Myr$ from the birth of the Solar System. Thus, just before the LHB, the size
distribution consisted of two parts. The objects smaller than hundreds of km had a size 
distribution set by their collisional evolution in the early massive EKB, whereas the larger 
objects retained a primordial distribution set by their formation process.
The LHB has then resulted in a dynamical depletion of the EKB, which was 
obviously size-independent \citep{Wyatt-et-al-2011}. As a result, the entire size distribution
must have been pushed down, retaining its shape.
During the LHB, the EKB has reduced its original mass by a factor of $\sim 1000$
\citep{Levison-et-al-2008}
and expanded to its present position. Both the reduction of mass and the increase of distance
to the Sun have drastically prolonged the collisional lifetime of the EKBOs
of any given size. As a result, during the subsequent $3.8\Gyr$ only the objects smaller than
about a hundred of meters in radius
(more accurate values will be obtained later, see Fig.~\ref{fig:lifetime})
experienced full collisional reprocessing.
We conclude that the size distribution in the EKB after the LHB, and in the present-day EKB,
is likely to consist of three parts.
Objects smaller than a hundred meters must currently reside in a 
collisonal equilibrium,
those with radii between a hundred meters and hundreds of kilometers inherit the collisional 
steady-state of the massive and compact belt of the pre-LHB stage, and the largest EKBOs still 
retain a primordial size distribution from their accretion phase.

\subsection{Setup of the collisional simulations}

To obtain the dust distributions in the present-day EKB, which is the goal of this paper,
we use our collisional code
{\it ACE} ({\it Analysis of Collisional Evolution})
\citep{Krivov-et-al-2000,Krivov-et-al-2005,Krivov-et-al-2006,%
Krivov-et-al-2008,Loehne-et-al-2008,Mueller-et-al-2009}.
{\it ACE} simulates evolution of orbiting and colliding solids,
using a mesh of sizes $s$, pericentric distances $\hat{q}$, and eccentricities $e$ of objects
as phase space variables.
It includes the effects of stellar gravity, direct radiation 
pressure, Poynting-Robertson force, stellar wind, and several collisional outcomes
(sticking, rebounding, cratering, and disruption), and collisional damping.

If we were able to set an initial size and orbital distribution of bodies
(i.e., the one after the completion of the LHB) in a reasonable way, we could
simply run the code over $3.8\Gyr$ to see which dust distribution it yields.
Setting the initial distribution at largest EKBOs, i.e. the third
of the three parts of the entire size distribution described above, is straightforward.
Since the distribution of these objects remains nearly unaltered since the LHB,
their initial distribution should be nearly the same as the current one.
Accordingly, we populate the {\it ACE} bins
with the debiased population of known EKBOs,
as described in \cite{Vitense-et-al-2010}.

\begin{figure}
  \begin{center}
  \includegraphics[width=0.50\textwidth]{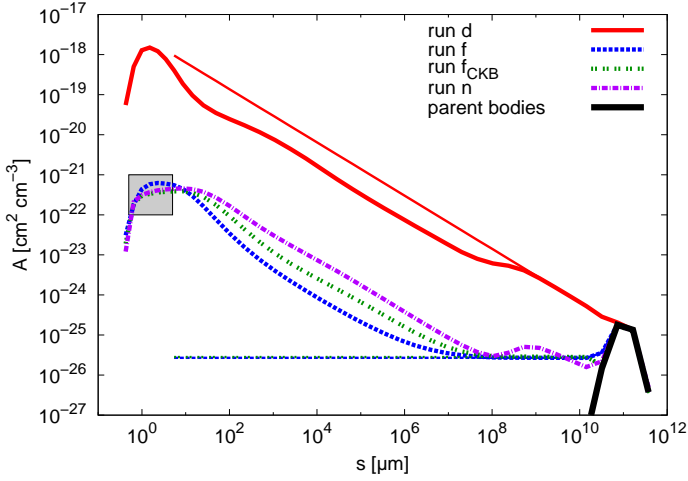}\\
  \end{center}
  \caption{
    Size distributions in different collisional runs.
    Thin lines are initial distributions, while thick ones correspond to an advanced
    state of collisional evolution.
    Note that the initial size distribution of the ``n'' run coincides with
    the debiased population of TNOs.
    $A$ is the cross-section density per size decade at a distance of $40\AU$.
    Note that $A={\rm const}$ (i.e. horizontal lines) correspond to a size distribution
    with $q=3$, where different-sized objects equally contribute to the cross-section.
    The gray shaded rectangle is a rough approximation of the particle dust flux given by New Horizons
    translated into the cross-section density and distances of the EKB.
    }
  \label{fig:size_dist}
\end{figure}

However, we do not know the second part of the distribution, at least
for objects between a hundred meters and $\sim 10\km$ where no or 
very few EKBOs have been discovered.
Given the lack of information on these objects,
we choose to extrapolate the size distribution to smaller objects
with a power law $\total N \propto s^{-q}\total s$.
The slope $q$ is unknown, so we explore the following possibilities
(thin lines in Fig.~\ref{fig:size_dist}):
\begin{enumerate}
 \item Run ``d'' (``Dohnanyi extrapolation'').
       We assume 
       the classical \citet{Dohnanyi-1969} law with $q=3.5$.
       This extrapolation is similar to the one used in \cite{Vitense-et-al-2010}.
 \item  Run ``f'' (``flat extrapolation'').
       We assume $q=3.0$ for $s<10\km$.
       Run ``f''can be treated as a rough proxy for
       a break in the size distribution at a few 
       tens of kilometers reported in the literature:
       $q \approx 1.9$ \citep{Fraser-Kavelaars-2009},
       $q \approx 2.0$ \citep{Fuentes-et-al-2009}
       and $q\approx 2.5$ \citep{Fuentes-Holman-2008}.
       Keeping in mind that the observed TNOs include several populations,
       and that the knowledge of scattered objects is particularly poor
       \citep{Vitense-et-al-2010}, we made an additional run
       ``f$_\text{CKB}$'' identical to ``f'',
       but without the scattered objects.
 \item Run ``n'' (``no extrapolation'').
       Here, we refrain from any extrapolation, assuming
       that the system was devoid of smaller objects initially.
       This formally corresponds to $q \rightarrow -\infty$.
\end{enumerate}

To complete specification of the initial conditions for {\it ACE} simulations,
we have yet to set the orbital distributions of 
the objects with $s\lesssim 10\km$.
On the absense of relevant observational data,
we simply assume that these objects inherit the pericentric distance
and the eccentricity from their parent bodies.
This means that, for every object that resides 
in a bin $\{s_i,\hat{q}_j,e_k\}$,
the bins $\{s_{l},\hat{q}_j,e_k\}$ are populated
with $(s_l/s_i)^{1-q}$ ($l<i$) objects (assuming logarithmic size bins).

As a minimum grain radius, we chose $0.4\mum$ and 
set size ratios of the adjacent bins of $1.5$ for dust sizes and
$2.3$ for the largest TNOs.
To cover the heliocentric distances from $4\AU$ to $400\AU$ we
used a logarithmically spaced pericenter grid with $21$ bins as well as a
linearly spaced eccentricity grid between $-1.5$ and $1.5$.
Note that negative eccentricities correspond to ``anomalous'' hyperbolic orbits,
which are open outward from the star and are attained by
smallest dust grains with a radiation pressure to gravity ratio $\;\beta>1$
\citep{Krivov-et-al-2006}.
To make sure that this, rather coarse, grid yields sufficiently accurate results,
we made another ``n'' run with a finer, more extended grid with a minimum grain radius
of $0.3\mum$ with size ratios of the adjacent bins of $1.25$ for dust sizes and
$1.58$ for the largest TNOs, $41$ pericenter bins and
eccentricity bins between $-5$ and $5$.
We found that our coarse grid leads to almost the same
results as the fine grid model.

As material, we assumed a mixture of $50\%$ ice \citep{Warren-1984} and $50\%$ 
astrosilicate \citep{Laor-Draine-1993} with a bulk density of $2.35\gperccm$.
The optical constants of the mixture were computed with the Bruggeman mixing rule
and the absorption coefficients with a standard Mie algorithm.
The values of other parameters, e.g. the critical fragmentation energy,
were the same as in \cite{Vitense-et-al-2010}.

\subsection{Results of the collisional simulations}

All the extrapolations described above are rather arbitrary, and the last one
obviously unrealistic.
A natural question is then, which of the models, and after which timestep,
should deliver the distributions that match the actual distributions of the EKB material
the best.
We start with the integration time.
Each of the runs was let to go as long as needed to reach a collisional equilibrium
at smaller sizes, but not too long in order to preserve the initial distribution
of larger objects. A boundary between ``smaller'' and ``larger'' sizes was arbitrarily set
to $s \sim 1 \km$.
We considered ``collisional equilibrium'' to have been reached, once the shape of
the size distribution stopped changing.
Note that, to meet these criteria in the ``n'' run, we had to let the
system evolve much longer than the age of the Universe. Of course,
this ``modeling time'' should not be misinterpreted as physical time of the EKB evolution.
This was simply the time needed for the population of large
bodies to generate sufficient amount of smaller debris down to dust sizes.

\begin{figure}
  \begin{center}
 \includegraphics[width=0.50\textwidth]{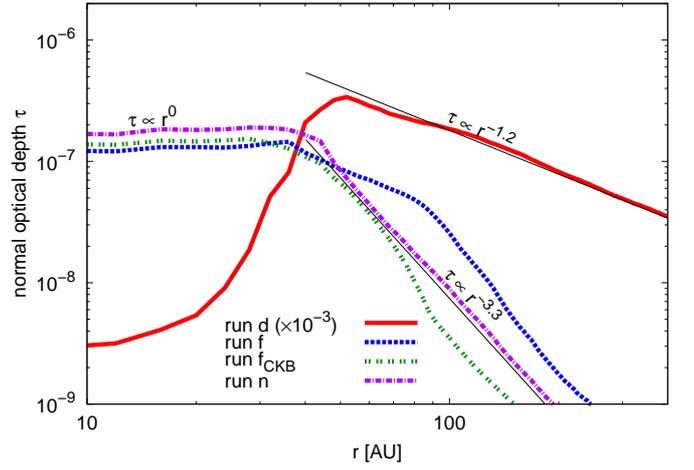}\\
  \end{center}
  \caption{
    Normal optical depth for the same {\it ACE} runs and time instants
    as in Fig.~\ref{fig:size_dist}.
  }
  \label{fig:tau}
\end{figure}

\begin{figure}
  \begin{center}
  \includegraphics[width=0.50\textwidth]{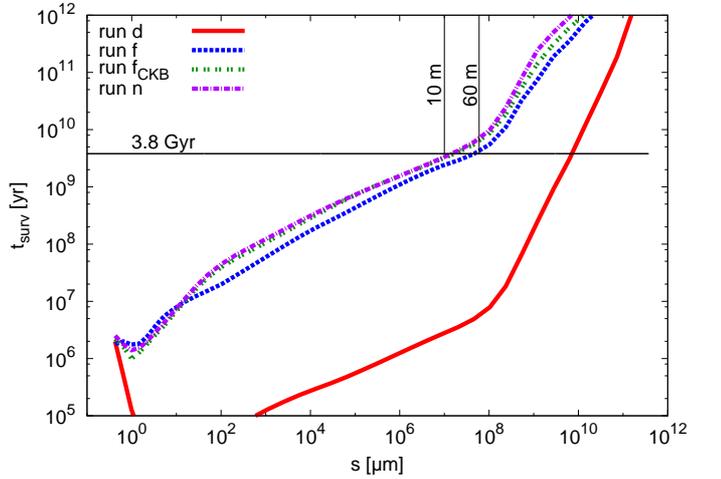}\\
  \end{center}
  \caption{
    Lifetimes of dust grains and parent bodies for the same {\it ACE} runs and
    time instants as in Fig.~\ref{fig:size_dist}.
    Particles below the $3.8\Gyr$ line are in a collisional equilibrium after the LHB.
    A steep rise in the lifetime at $s \sim 300\m$ corresponds to the strength--gravity
    transition of the critical disruption energy.
  }
  \label{fig:lifetime}
\end{figure}

The results obtained over the integration interval chosen in this way are
shown in Figs.~\ref{fig:size_dist}--\ref{fig:lifetime} with thick lines.
These three figures show the size distribution,
the radial profile of the normal gemetrical optical depth,
and the collisional lifetime of the objects, respectively.
We note that at an earlier stage of evolution the cross-section density
and the normal optical depth would be lower, and the lifetime of dust grains
longer, while a later stage of evolution would lead to more dust and
therefore to a higher cross-section density and optical depth 
and reduced lifetime of the particles.

But which of the models, ``d'', ``f'', of ``n''~--- if any~---
matches the actual dust distribution in the present-day EKB the best?
The only way to answer this question is to compute, for each of the simulations,
the observables and compare them with in-situ spacecraft measurements and thermal emission
constraints.
Although an in-depth analysis of the data is deferred to Sec.~\ref{sec:in-situ},
we now take a first quick look.
The gray shaded rectangle in Fig.~\ref{fig:size_dist}
is a rough approximation of the dust flux data collected by New Horizons, translated into
the cross-section density and extrapolated to the distance of the classical EKB.
A comparison with the evolved curves demonstrates
that the ``d'' run is far too dusty.
It cannot reach an evolutionary stage which would be consistent with the
measurements (and with the upper limit from the non-detection of the thermal emission).
Thus we conclude that a straightforward extrapolation
from debiased EKBOs to dust sizes has to be ruled out.
Therefore we have shown here, with a compeletely different type of argument,
that a break in the size distribution {\it has} to be 
present in the EKB, as found from the analysis of TNO observations
\citep{Fuentes-Holman-2008,Fraser-Kavelaars-2009,Fuentes-et-al-2009}.

How about the other runs? Both the ``f'' and the ``n'' runs are consistent with
the observational data; we will confirm this in Sec.~\ref{sec:in-situ}
by a more thorough analysis.
Thus~--- unfortunately~--- we cannot constrain the size
distribution of EKBOs more tightly. Nor can we say which of the dust distributions,
the one of the ``f'' run or the ``n'' run, can be expected in the EKB,
although the shape of the curves in these runs is different.
(The only common feature shared by all the curves is an abrupt drop
at $\approx 0.5\mum$, which is the limit, below which the grains are swiftly
removed from the system by radiation pressure.)
We now come to an analysis of these differences.

The size distribution in the ``d'' run, which we rejected due to violation
of observational constraints, is typical of a collision-dominated disk.
At all sizes, the dust transport is less efficient than the collisional grinding,
and the cross-section density peaks just above the blowout limit
\citep{Krivov-et-al-2006,Thebault-Augereau-2007}.
This is also confirmed by the radial profile shown in Fig.~\ref{fig:tau}.
The outer slope of $\approx 1.2$ is in a good agreement with an approximate analytic solution
for a collision-dominated disk that predicts a slope of $\approx 1.5$
\citep{Strubbe-Chiang-2006}.
Also, there is a clear decrease of the optical depth toward the star,
caused by collisional elimination of the particles.
Since we can rule out this extrapolation
further results for the ``d'' run
are presented but not discussed anymore.
As mentioned above, the ``d'' run is essentially the same as the 
run in \cite{Vitense-et-al-2010} (made for the expected EKB, with the Poynting-Robertson 
effect included), so for a detailed analysis of the ``d'' run we refer to our previous work.
As shown here, this run fails to describe the actual present-day EKB
in the Solar System.
Nevertheless, the results and conclusions presented in \cite{Vitense-et-al-2010}
would still be valid for an EKB analog, in which all objects down to kilometer
in size are in collisional equilibrium.

The size distribution in the ``n'' run is different.
It shows a broad maximum at $\sim 100\mum$, which indicates that
particles smaller than that are transported inward from the dense part of the disk
before they are lost to collisions \citep{Wyatt-et-al-2011}.
The inner part of the radial profile in Fig.~\ref{fig:tau} is nearly constant,
and the outer one reveals a steeper slope
of $\approx 3.0$, as predicted analytically for
a transport-dominated disk ($\approx 2.5$,
\citeauthor{Strubbe-Chiang-2006} \citeyear{Strubbe-Chiang-2006}).
Note that the outer profiles are generated by particles in a narrow range of sizes
around the blowout limit. The coarse size grid in our models therefore limits the accuracy 
with which we can reproduce these slopes.

The ``f'' run seems to be intermediate.
Although the maximum in the size distribution is broader than in the ``d'' run,
it still resembles the curves typical of collision-dominated disks.
However, the profile of the normal optical depth (Fig.~\ref{fig:tau}) stays nearly constant
inside the main belt, which is typical of transport-dominated disks \citep[e.g.][]{Wyatt-2005}.

Figures~\ref{fig:size_dist}--\ref{fig:lifetime}
also present the results of the additional ``f$_\text{CKB}$'' run,
from which we excluded scattered objects as dust parent bodies.
Figure~\ref{fig:size_dist} shows, somewhat unexpectedly, that the results
of ``f'' and ``f$_\text{CKB}$'' runs differ from each other: the dust disk
in the latter turns out to be transport-dominated, similar to the ``n'' run.
The question is why.
This is not because dropping the scattered objects just reduces the amount of material
in the EKB, resulting in reduced collisional rates.
A test simulation, in which we 
artificially augmented the mass of the classical
EKB to the total mass of the expected EKB, 
brought qualitatively the 
same results as the ``f$_\text{CKB}$'' run.
Instead, the answer can be found in the method of extrapolation.
As explained before, we filled
the $(s,q,e)$-bins with our debiased population of EKBOs and extrapolated towards smaller sizes
with a power law into the same $(q,e)$-bins.
That means that we transfer the high eccentricities of the large scattered
objects to all smaller ones.
Although higher eccentricities do not lead to higher collisional rates 
\citep[see][discussion after their Eq. (17)]{Krivov-et-al-2007},
they increase the relative velocities, making collisions more disruptive.
In the ``f'' run a large amount of $s<10\mum$ particles is produced, 
leading to a higher number and cross-section density for these particles,
which in turn leads to a higher collisional rate and 
a shorter collisional lifetime for larger particles (Fig.~\ref{fig:lifetime}).
Without the eccentric orbits of scattered objects (``f$_\text{CKB}$'' run),
the relative velocities are moderate, collisions are less disruptive and fewer small particles
are produced.
Therefore, destruction of larger grains becomes less efficient,
leading to a prolonged collisional lifetime.

The above discussion demonstrates that it remains unclear whether
the EKB dust disk is transport- or collision-dominated.
It is most likely that it is either transport-dominated
or intermediate between a collision- and transport-dominated disk.
However, in all the runs considered, the inner part of the dust disk (inside the
classical EKB) has a nearly contant radial profile of the optical
depth of $\tau_{\perp} \sim 1 \times 10^{-7}$ (Fig.~\ref{fig:tau}).
(For comparison, the in-plane optical depth for $r>10\AU$ 
is $\tau_{\parallel}=1\dots2\times 10^{-6}$.)
This suggests that collisions in the inner part of the disk can be neglected.
This justifies that in this section we first simulated a completely planet- and
sublimation-free EKB and will include the effects of planetary scattering and
sublimation later, in Sec.~\ref{sec:planets} and \ref{sec:sublimation}.

Figure~\ref{fig:lifetime} shows the mean collisional lifetimes averaged over {\it all} distances
for the same {\it ACE} runs at the same time instants.
Note that the collisional lifetime in the main belt is much shorter than the average one
because the density there is much higher and therefore collisions are more frequent.
The horizontal line represents a lifetime of $3.8\Gyr$ which is the time elapsed after the LHB.
All grains below this line are in a collisional equilibrium in the present EKB.
For all simulations this size is just about $(10\dots 60)\m$.
The distribution of all objects larger than that equilibrium size was set before the LHB
and cannot be constrained with our collisional model.

\section{Influence of planets}\label{sec:planets}

Giant planets interact gravitationally with dust in the outer Solar System.
On the one hand,
the grains drifting inward by the Poynting-Robertson (P-R) drag \citep{Burns-et-al-1979}
can be captured by planets into outer mean-motion resonances
\citep[e.g.][]{Liou-Zook-1999,MoroMartin-Malhotra-2002,MoroMartin-Malhotra-2003,MoroMartin-Malhotra-2005,Kuchner-Stark-2010}.
On the other hand, the grains that cross the planet's orbit can be scattered.
Both effects are able to modify the size and spatial distibution of dust in the disk.
In this section, we investigate the efficiency of capturing and scattering.

\subsection{Resonant trapping}\label{sec:resonance}

\citet{Mustill-Wyatt-2011} developed a general formalism to calculate the capture probability
of a particle into the first- and second-order resonances with a planet.
Their theory is valid for any convergent differential
migration of the particle and the planet
(for instance, if the particle is drifting inward and the planet is migrating outward).
Their results are presented in terms of 
the generalized momentum $J$
and a dimensionless drift rate $\dot{B}$
($\dot{\beta}$ in their paper).

The generalized momentum is related to the orbital eccentricity of the
particle reaching the resonance location, $e$, while
the dimensionless drift rate $\dot{B}$  can be expressed through the
differential change rate of the particle's semimajor axis, $\dot{a}_\text{res}$,
and the semimajor axis itself, $a_\text{res}$.
In what follows, we make estimates for the $3:2$ resonance with Neptune.
Using Eqs.~(3) and (4) of \citet{Mustill-Wyatt-2011},
we find the following conversion relations:
\begin{align}
 J &= 5893.36 \left(\frac{m_\text{N}}{m_\oplus}\right)^{-2/3}e^2
 \label{conv1}
 \\
 \dot{B} &= -0.818921\left(\frac{m_\text{N}}{m_\oplus}\right)^{5/6}
 \sqrt{\frac{a_\text{N}}{\AU}}\frac{a_\text{N}}{a_\text{res}}\frac{\dot{a}_\text{res}}{1\AU\Myr^{-1}},
 \label{B}
 \end{align}
where $m_{\oplus}$ and $m_\text{N}$ denote the masses of Earth and Neptune, respectively,
and $a_\text{N}$ is the semimajor axis of the Neptune orbit.

We now assume that $\dot{a}_\text{res}$ is caused by P-R drag
\citep{Wyatt-Whipple-1950}
\begin{align}
 \dot{a}_\text{res}
 &=
 - 1.3\frac{\beta}{c} \frac{G M_{\odot}}{a_\text{res}} \frac{2 + 3 e^2}{(1-e^2)^{3/2}}
\nonumber\\
 &=
 -815\frac{\beta}{a_\text{res}[\text{AU}]}\frac{2 + 3 e^2}{(1-e^2)^{3/2}}\frac{\AU}{\Myr} ,
 \label{dot a_res}
\end{align}
where the prefactor $1.3$ accounts for the enhancement of P-R drag
by solar wind drag \citep{Burns-et-al-1979} and $\beta$ is the radiation pressure to
gravity ratio for the particle.
The enhancement by solar wind drag is included in all further analysis
but we will call it just P-R drag for brevity.
The $\beta$-ratio not only controls the drift rate, it also
reduces the effective solar mass felt by the particle by a factor of $(1-\beta)$.
This affects the resonance location, so that
$a_\text{res}$ reads:
\begin{equation}
 a_\text{res} = a_\text{N}\sqrt[3]{1-\beta} \left(3/2\right)^{2/3}.
 \label{a_res}
\end{equation}
Inserting Eqs.~(\ref{dot a_res}) and (\ref{a_res}) into Eq.~(\ref{B}),
the latter takes the form
\begin{equation}
 \dot{B} = 1.6\frac{\beta}{(1-\beta)^{2/3}}\frac{2+3e^2}{(1-e^2)^{3/2}}.
 \label{conv2}
\end{equation}

\begin{figure}
  \begin{center}
 \includegraphics[width=0.50\textwidth]{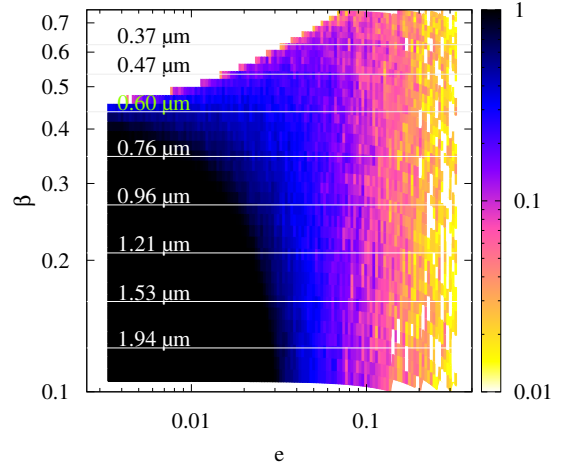}\\
  \end{center}
  \caption{
    Capture probability of a single particle with given $\beta$ and $e$ at the location of the 3:2 resonance with Neptune.
  }
  \label{fig:cap-prob}
\end{figure}

Using the capture probabilities as functions of $J$ and $\dot{B}$ 
from \citet{Mustill-Wyatt-2011} and applying Eqs.~(\ref{conv1}) and (\ref{conv2}),
we computed the probabilities as functions of $e$ and $\beta$ (or equivalently,
particle radius $s$).
The results for the 3:2 resonance with Neptune
are shown in Fig.~\ref{fig:cap-prob}.
Although capturing for grains $s>0.6\mum$ and $e<0.03$ seems unavoidable,
it is not obvious, what is the fraction of particles of those sizes 
that will actually have such low eccentricities.
The reason is that small particles, when released from parent bodies
in nearly-circular orbits, are sent by radiation pressure into large
and highly-eccentric orbits.
Subsequently, drag forces reduce the semimajor axes and eccentricities of the grains.
Yet, it is not clear how low the eccentricities will be by the time when the
grains will have reached the resonance location.

To find this out, we first consider
parent bodies with elements $a_\text{p}$ and $e_\text{p}$
and compute the initial semimajor axis $a_\text{i}$ and
the eccentricity $e_\text{i}$ of a grain upon release.
To this end, we use Eqs. (19)--(20) of \citet{Krivov-et-al-2006},
in which we neglect the mass of the projectile
compared to the mass of the target, i.e. the  parent body,
and assume that ejection occurs at the pericenter of the parent body orbit:
\begin{align}
 a_i &= a_\text{p}\frac{(1-\beta)(1-e_\text{p})}{1-e_\text{p}-2\beta}\label{eq:burns1}\\
 e_i &= \frac{\beta+e_\text{p}}{1-\beta}.\label{eq:burns2}
\end{align}
Subsequently, the P-R drag will decrease $a_i$ and $e_i$.
Denoting by $e_f$ the final eccentricity~-- i.e. the one the grain will have
at the location of a resonance, $a_\text{res} = a_f$~--
and using the dependence of $\dot{a}$ on $\dot{e}$ as given in \cite{Wyatt-Whipple-1950}
\begin{equation}
 \frac{\total e}{\total a} = \frac{5}{2a}\frac{e(1-e^2)}{2+3e^2}
\end{equation}
leads to
\begin{equation}
 \left(\frac{e_f}{e_i}\right)^{4/5}\frac{1+e_i^2}{1-e_f^2} = \frac{a_f}{a_i}.
\end{equation}

As an example, a plutino with $a_\text{p} \approx 39\AU$ and
$e_\text{p} = 0.1$ will release a $\beta = 0.3$ particle into an orbit
with $a_i = 82\AU$ and $e_i = 0.57$.
The $3:2$ resonance with Neptune for this particle is located at
$a_\text{res} \approx 35\AU$. At that location, the grain will have
$e_f = 0.13$.

\begin{figure*}
  \begin{center}
 \includegraphics[width=0.90\textwidth]{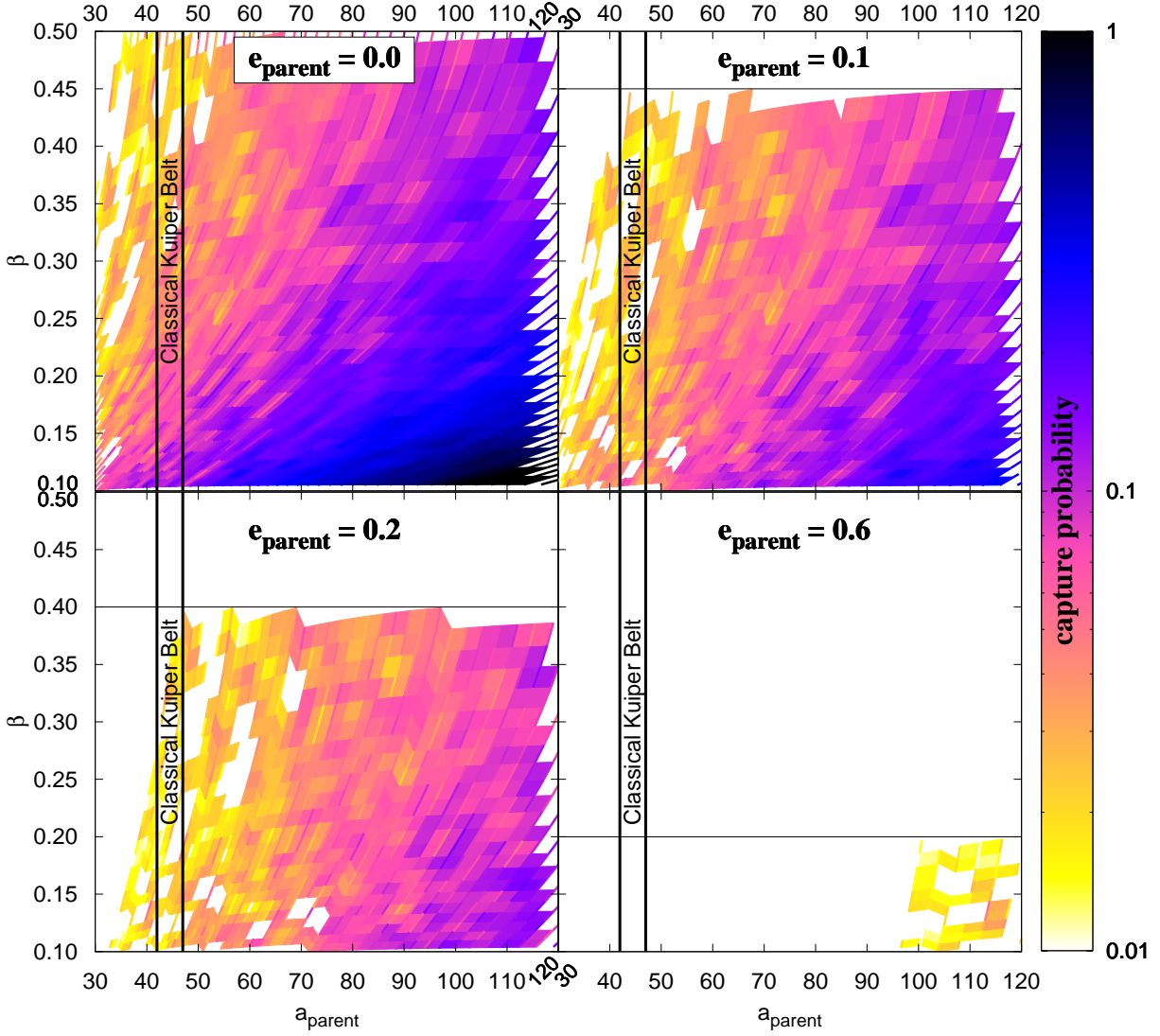}\\
  \end{center}
  \caption{
    Capture probability of a $\beta$ particle released by a parent body for the 3:2 resonance with Neptune.
    $e_\text{p} = 0.6$ shall represent the population of scattered objects.
    Note that all particles above the horizontal lines have initial eccentricites of $e>1$
    and will be removed from the system (Eq.~\ref{eq:burns2}).
  }
  \label{fig:cap-prob-parent}
\end{figure*}

With Eqs. (6) -- (9) and the data of \cite{Mustill-Wyatt-2011}
it is possible to calculate the capture probablility for each resonance
and particle size for given $a_\text{p}$ and $e_\text{p}$.
As a word of caution, we note that the actual dust dynamics can be more 
complicated. One complication
is that the initial semimajor axis (Eq.~\ref{eq:burns1}) for sufficiently
small particles is often so large that the grain has to pass several other resonances
before it reaches the $3:2$ one. At these resonances, particles with high migration rates
and small eccentricities will experience an eccentricity jump when they are not captured.
As a result, our model will underestimate the final eccentricity at the $3:2$ resonance
and so overestimate the capture probability.
Slow migration rates and small eccentricities will result in the opposite
effect~--- an eccentricity decrease and a probability increase~--- so an
underestimation of the capture probability is also possible \citep{Mustill-Wyatt-2011}.
A detailed modeling of this problem is beyond the scope of this paper.

Fig.~\ref{fig:cap-prob-parent} shows the probability
of capture into the 3:2 resonance with Neptune.
The probability is the highest for $e_p=0$,
but even in that case it does not exceed 
$\approx 20\%$ for dust grains below $2\mum$ when released from classical EKBOs.
Increasing the eccentricity of the EKBOs {and decreasing the grain size
reduces} the  capturing efficiency.
For $e_\text{p} = 0.6$, which can be considered representative for scattered
disk objects, the capturing probability is 
just a few per cent. 
For $e_\text{p} = 0.1$ (typical classical EKBOs) and $s\sim 1\mum$
(just above the threshold of the New Horizons dust detector),
the trapping probability is still below $10\%$.
Given these results, resonant capturing can be considered unimportant
for the purposes of this paper and will be neglected.

\subsection{Gravitational scattering}\label{sec:scattering}

Since P-R drag continuously decreases the particle's distance from the Sun,
the grain will eventually reach the orbit of a planet.
As this happens, the grain can either fall onto the planet,
be scattered, or pass the planet without interaction.
In the first two cases the particle will be lost.
To determine the surviving fraction we used a numerical code that
calculates the orbital evolution of a single particle, taking into account
the gravity of one planet and the P-R effect.
For each $\beta$-value
listed in Table~\ref{tab:surv-rate} (these are the same values
as used in our collisional simulations) we started $10.000$ particles,
with an EKB-like $a,e,i$ distribution taken from the upper panel of Fig.~5 from
\cite{Vitense-et-al-2010}.
A particle was counted as a survivor as soon as its apocentric distance
became smaller than the pericentric distance of the planet.

The results are listed in Table~\ref{tab:surv-rate}
for Neptune, Uranus and Saturn.
As expected, the surviving rate decreases for larger grains with lower
migration rates.
For Neptune and Uranus the ejection rate
is negligible and will not alter the dust flux significantly
(Sec.~\ref{sec:in-situ}).
However, Saturn ejects nearly a half of the dust grains.
As we will see in Sec.~\ref{sec:in-situ}, Saturn's influence
is important to explain the in-situ measurements, but
all three planets have little effect on the thermal emission of the
EKB dust (Sec.~\ref{sec:SED}).

\begin{table}
\caption{
  $\beta$ values and corresponding sizes, masses and surviving rates for particles passing
  Neptune, Uranus and Saturn.
  Particles between $10^{-12}\g < m < 10^{-9}\g$ can be measured by
  the New Horizons dust counter \citep{Horanyi-et-al-2008}.
  }
\label{tab:surv-rate}
\centering
\begin{tabular}{c c c c c c }
\hline\hline
 $\beta$ & $s$ [\textmu{}m] & $m$ [g] & $\neptune_\text{surv}$ & $\uranus_\text{surv}$ & 
$\saturn_\text{surv}$\\
\hline
$0.404$ &       $0.65$ & $2.7\times 10^{-12}$ & 96.8\% & 97.4\% & 79.7\%\\
$0.259$ &       $0.99$ & $9.5\times 10^{-12}$ & 93.8\% & 95.1\% & 66.0\%\\
$0.164$ &       $1.5$  & $3.4\times 10^{-11}$ & 88.7\% & 90.0\% & 57.0\%\\
$0.106$ &       $2.3$  & $1.2\times 10^{-10}$ & 82.2\% & 80.1\% & 50.3\%\\
$0.070$ &       $3.5$  & $4.3\times 10^{-10}$ & 78.3\% & 73.9\% & 47.1\%\\
\hline
\end{tabular}
\end{table}

As shown in the previous section, the EKB dust disk is transport-dominated for
small particles, which means that collisions play a minor role.
Therefore, gravitational scattering can simply be implemented by
multiplying the distribution obtained in the collisional simulation
by the surviving rates for the corresponding particle sizes and distances.

\section{Influence of sublimation}\label{sec:sublimation}

\begin{figure}
  \begin{center}
  \includegraphics[width=0.50\textwidth]{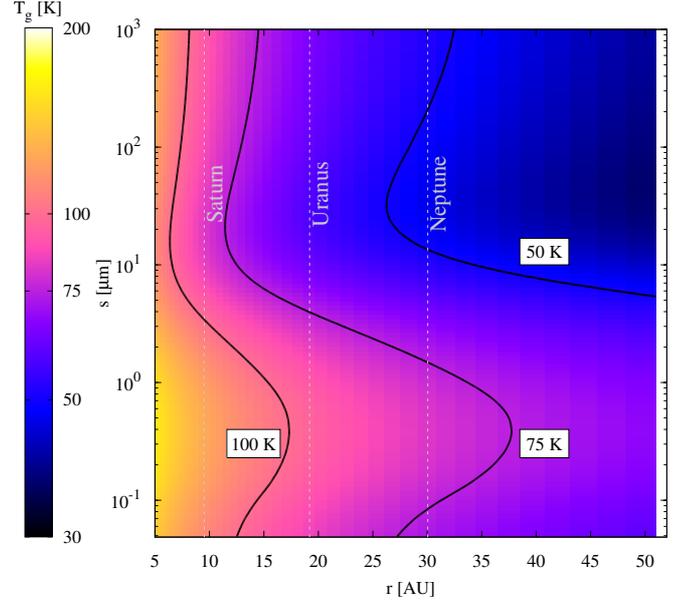}\\
  \end{center}
  \caption{
    Temperatures of the dirty-ice particles for different distances.
    The solid lines correspond to $100\Kelvin$, $75\Kelvin$ and $50\Kelvin$.
    Sublimation occurs at $T_\text{subl} = 100\Kelvin$.
    Sublimation distance increases with decreasing size because
    the emission efficiency of small grains is lower, which makes them hotter.
  }
  \label{fig:temp_function}
\end{figure}

When drifting inward, dust grains will not only suffer interaction with planets,
but they will also be heated up due to the decreasing distance to the Sun.
Our dust particles are composed of ''dirty ice``
($50\%$ ice and $50\%$ astrosilicate in volume).
Their icy part sublimates at $\approx 100\Kelvin$
\citep{Kobayashi-et-al-2008,Kobayashi-et-al-2009,Kobayashi-et-al-2011}.
Since the EKB dust disk is radially optically thin,
the temperature of a dust grain is determined by the energy balance
between the absorption of incident solar radiation and the thermal emission   
of the grain. We neglect the latent heat of
sublimation because its contribution is minor
\citep{Kobayashi-et-al-2008}.
The sublimation distance $r_\text{subl}$, where the temperature of a
particle reaches $100\Kelvin$, depends on its size.
If the particles are larger than $\lambda / (2\pi)$, where
$\lambda$ is the peak wavelength of emission,
the absorption and emission cross-sections are
approximately the same as the geometrical one,
and these particles can be assumed to be blackbody radiators.
Since for $T = 100\Kelvin$ the maximum is at $\lambda\sim 30\mum$,
this is the case for grains with $s > 5\mum$.
Temperatures of smaller particles are obtained
by solving the thermal balance equation
\citep[see, e.g.,][]{Krivov-et-al-2008}.
Fig.~\ref{fig:temp_function} shows the resulting temperatures for different sizes
and distances, with three isotherms overplotted. The leftmost one corresponds
to $100\Kelvin$.
Empirically we can approximate the dependence of the sublimation distance (in AU)
on the size (in micrometers) by:
\begin{equation}
r_\text{subl} =
\begin{cases}
-10.2\sin(0.26s)+16.85 & s\leq 5.0\mum \\
\qquad\qquad\qquad  8.0& s > 5.0\mum
\end{cases}
\end{equation}
Note that this function does not have a physical meaning and is only needed to
implement sublimation into our model.

The outcome of sublimation depends on the structure of icy grains.
If a single icy particle is an aggregate of small grains, each having
$\beta \ga 0.5$, the resulting grains will be blown out and therefore
no grains should be present inside $r_\text{subl}$.
However, if the constituent monomers have $\beta \la 0.5$, the number
density of grains inside $r_\text{subl}$ will increase.
Since both is inconsistent with the dust flux
measured by spacecraft, a single icy grain is likely to contain a single
core of refractory material covered with an ice mantle \citep{Kobayashi-et-al-2010}.
For our dirty-ice grains sublimation will result in a
$100\%$ silicate particle which
has a half of the volume of the original particle.
The radius of the resulting particle is simply
\begin{equation}
  s_\text{silicate} =  \sqrt[3]{0.5}s_\text{icy},
\end{equation}
and the mass is given by
\begin{equation}
 m_\text{silicate} = 0.5\frac{\rho_\text{silicate}}{\rho_\text{icy}}m_\text{icy},
\end{equation}
with $\rho_\text{icy} = 2.35\gperccm$ being the bulk density of the dirty ice and
$\rho_\text{silicate} = 3.35\gperccm$ of the astrosilicate.
The typical sizes and $\beta$-values of the particles before and after
sublimation, together with their sublimation distances,
are given in Table~\ref{tab:subl}.

We now discuss how sublimation affects the distribution of dust.
The particles born through collisions in the
Edgeworth-Kuiper belt have eccentricities roughly comparable to their $\beta$ values (Eq.~\ref{eq:burns2}).
Although damped by P-R drag, their eccentricities in the sublimation zone
are typically larger than $0.05$.
Particles with $e>0.05$ will experience a
rapid sublimation without pile-up and dust ring
formation \citep{Kobayashi-et-al-2009};
\citep[cf.][their Fig.~8]{Burns-et-al-1979}.
Next, although sublimation in our model does not eliminate the particles and thus
preserves their number, it reduces their spatial number density.
This is because the number density of particles is  
inversely proportional to their drift rates in the steady state.
Because $\dot{a}\propto \beta$,
the increase of $\beta$ due to sublimation lessens the number density of
particles.
Based on Table~\ref{tab:subl},
the change is estimated to be only about
$20\%$, see Fig.~\ref{fig:in-situ} below.
However, with in-situ dust detectors measuring only grains above a certain threshold,
the observable dust flux decreases more strongly.

If we assume that the orbital changes due to the
change in size and therefore changing interaction with the stellar radiation
are small, we can implement sublimation into our collisional
results the same way as gravitational scattering by
simply correcting sizes and cross-section- and mass density
for the affected bins.
Since planetary scattering and sublimation are independent processes,
the order of implementation does not matter.

\begin{table}
\caption{
  Sizes and $\beta$-values before and after sublimation
  and the corresponding sublimation distances; for particles larger than
  $5.0\mum$ blackbody temperatures are assumed.
  }
\label{tab:subl}
\centering
\begin{tabular}{c c c c c}
\hline\hline
 $s_\text{icy}$ [\textmu{}m] & $\beta_\text{icy}$ & $r_\text{subl}$ [AU] & $s_\text{silicate}$ [\textmu{}m] & $\beta_\text{silicate}$\\
\hline
$0.425$  & $0.576$ & $15.7$    & $0.337$ & $0.652$\\
$0.648$  & $0.404$ & $15.1$    & $0.514$ & $0.428$\\
$0.989$  & $0.259$ & $14.3$    & $0.785$ & $0.280$\\
$1.51$   & $0.164$ & $13.0$    & $1.20$ & $0.184$\\
$2.30$   & $0.070$ & $11.1$    & $1.83$ & $0.121$\\
$3.51$   & $0.046$ & $\;\;8.8$ & $2.79$ & $0.079$\\
$>5.0$  & & $\;\;8.0$& $\sqrt[3]{0.5}s_\text{icy}$ &\\
\hline
\end{tabular}
\end{table}

\section{Comparison with spacecraft measurements}\label{sec:in-situ}

\begin{figure}
\includegraphics[width=0.50\textwidth]{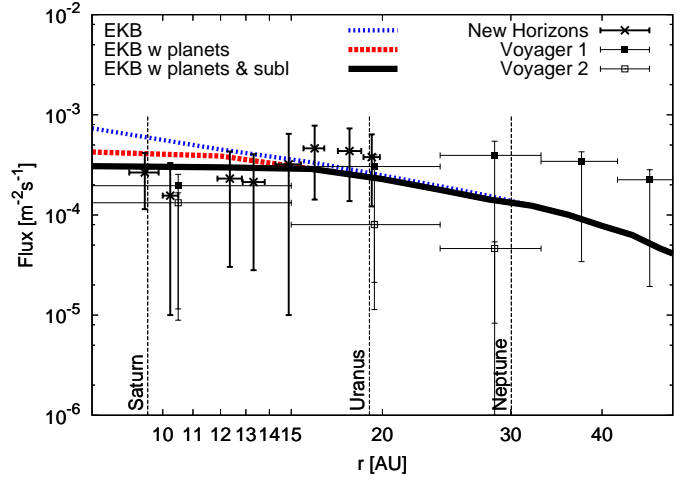}\\
\caption{Simulated particle flux
in comparison with in-situ measurements by New 
Horizons and Voyager $1$ and $2$.
The solid black line takes into account planets and sublimation and
represents our best fit to the data 
taken from \cite{Poppe-et-al-2010}, \cite{Han-et-al-2011} and \cite{Gurnett-et-al-1997}.
The dotted and dashed lines show the dust flux for the unperturbed EKB 
and after planetary scattering without sublimation, respectively.
}
  \label{fig:in-situ}
\end{figure}

The Student Dust Counter on-board the New Horizons spacecraft is capable
of detecting impacts
of grains with $10^{-12}\g<m<10^{-9}\g$ and can distinguish grain masses apart by a factor of $2$ 
between  $0.5\mum<s<5\mum$ \citep{Horanyi-et-al-2008}.
The first results from \cite{Poppe-et-al-2010} indicate particle fluxes up to $1.56\times10^{-4}\persmpers$.
The results of \cite{Han-et-al-2011} show a slight increase of the flux for $r>15\AU$.
The particle flux can be calculated via
\begin{equation}
 F^\text{p}_{\text{dust}} = \int mnv_\text{rel} \total\, (\ln m)
\end{equation}
with $m$ being the mass of the particle, $n$ the number density per logarithmic mass and
$v_\text{rel}$ the relative velocity between the spacecraft and the particle.
The first two values are a direct output of our simulation.
The relative velocity was assumed to be $v_\text{rel} = 15.54\kmpers$,
according to the official New Horizons web 
page\footnote{\textit{http://pluto.jhuapl.edu/mission/whereis\_nh.php}
(Last accessed on 2~September 2011)}.
Based on the results of Secs.~\ref{sec:dust_production}--\ref{sec:sublimation},
we calculated the dust fluxes for the EKB dust disk
unaffected by planets and sublimation, the one with planets and the one with 
planets and sublimation.
Although the separate contributions of planets and sublimation are rather 
low, their combination can alter the dust flux up to a factor of three, whereby Saturn plays
the most important role.
In Fig.~\ref{fig:in-situ} the results of run ``f'' are shown.
The right evolutionary state of the simulation (i.e., timestep) was chosen in the 
following way.
As seen in Fig.~\ref{fig:in-situ}, the black solid line can be assumed to
be a constant for $r<20\AU$. Taking this assumption we fitted the New Horizons data
from \cite{Poppe-et-al-2010} and \cite{Han-et-al-2011} by a
constant line to $F^\text{p}_{\text{dust}} \approx 3\times 10^{-4}\persmpers$ 
and searched for the timestep which agrees with the model the best.

According to \cite{Gurnett-et-al-1997},
the Voyager $1$ and $2$ plasma wave instruments, that acted as ``chance'' dust detectors,
have a mass threshold of $m>1.2\times 10^{-11}\g$,
which is one order of magnitude higher than for the New Horizons dust counter.
Accordingly, we
rescaled the Voyager data to the New Horizons threshold, 
with a power law slope of $q=-1$ obtained 
in our simulation for the corresponding masses (Fig.~\ref{fig:size_dist}) at $40\AU$.
Since the instruments aboard Voyager I and II were neither designed to detect 
dust impacts nor calibrated for this purpose and traversed the outer Solar System in
highly inclined orbits, their dust measurements should be compared with our model
with great caution.

Simulations f, f$_\text{CKB}$, and n were treated the same way.
Since for the particle sizes in question
($s<5\mum$) all modeled disks are transport-dominated, the results
do not differ much from each other and lead approximately
to the same fits as for the ``f'' run.
Therefore these results are not shown in Fig.~\ref{fig:in-situ}.

\section{Thermal emission constraints}\label{sec:SED}

To calculate thermal emission of dust in the EKB,
we computed the photospheric spectrum of the Sun using the NextGen models
\citep{Hauschildt-et-al-1999}.
The equilibrium dust temperatures were obtained 
by the procedure of \cite{Krivov-et-al-2008}.
As in the collisional simulations,
we adopted the ``dirty ice'' consisting of equal 
volume fractions of ice \citep{Warren-1984} and astrosilicate \citep{Laor-Draine-1993}.
As explained in Sec.~\ref{sec:sublimation} the sublimation distance
depends on the particle size. Therefore we divided the EKB
into $6$ sub-rings to handle the different emission properties of
the dirty ice and pure astrosilicate (Table~\ref{tab:subl_2}).

\begin{table}
\caption{
  The EKB divided into $6$ sub-rings: material composition together with the sizes 
  and distances for which that composition was adopted.
  }
\label{tab:subl_2}
\centering
\begin{tabular}{c c l l}
\hline\hline
ring \# & material & $s$ [\textmu{}m] & distance [AU]\\
\hline
1   	& astrosilicate & $0.425<s<\infty$      & $\;\;0 < r < \;\;8$\\
2   	& dirty ice	& $3.51\;\;<s<\infty$   & $\;\;8\leq r < 12$\\
3   	& astrosilicate & $0.425<s\leq3.51$     & $\;\;8\leq r < 12$\\
4   	& dirty ice	& $1.51\;\;<s<\infty$   & $12\leq r < 16$\\
5   	& astrosilicate & $0.425<s\leq1.51$     & $12\leq r < 16$\\
6   	& dirty ice	& $0.425<s<\infty$      & $16\leq r < \infty$\\
\hline
\end{tabular}
\end{table}

\begin{figure}
  \begin{center}
  \includegraphics[width=0.50\textwidth]{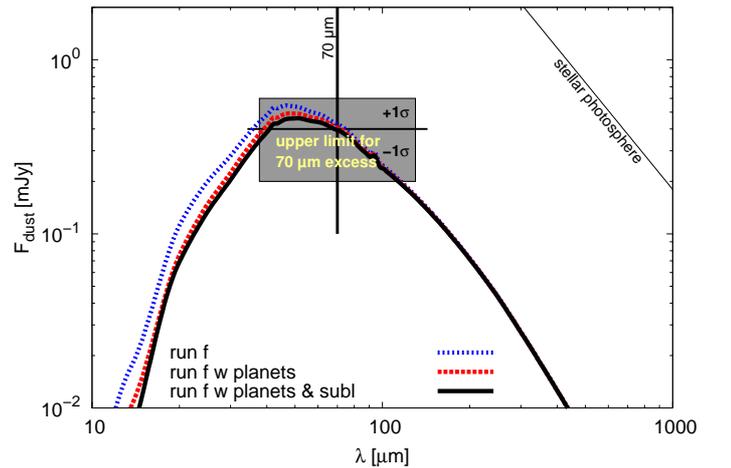}\\
  \end{center}
  \caption{ Spectral energy distribution of the EKB
            including planets and sublimation (solid black line),
            without sublimation (dashed red line),
            and for an unperturbed EKB (dottted blue line).
            All the curves are based on the same run (``f'') and the same time 
            instant as in Fig.~\ref{fig:in-situ}.
          }
  \label{fig:seduce}
\end{figure}

To place our EKB in the context of extrasolar debris disks, we now consider the
EKB dust disk, as if it were viewed from outside.
The final spectral energy distribution (SED)
of the EKB dust disk, as seen from a reference distance of $10\pc$,
is shown in Fig.~\ref{fig:seduce}. 
The influence of planets and sublimation does not alter the shape,
peak position, and height significantly.
The SED, corrected for planets and sublimation,
peaks at $40-50\mum$ with a maximum thermal emission flux of $\approx 0.5\mJy$, which amounts 
to $\approx 0.5\%$ of the photospheric flux at that wavelength.
The predicted flux drops to $\approx 0.4\mJy$ at $70\mum$
and to $\approx 0.2\mJy$ at $100\mum$.
The fractional luminosity of the modeled EKB dust disk,
after applying corrections for planets and sublimation,
is $f_d = 1.2\times 10^{-7}$.

Our results are consistent with the upper limit, placed
by non-detection of the EKB dust emission at $70\mum$ with the COBE spacecraft.
That limit amounts to $1\pm 0.5\%$ of the solar photospheric flux \citep{Greaves-Wyatt-2010}
and is shown by the dark gray box in Fig.~\ref{fig:seduce}.

Would EKB analogs around nearby stars be detectable, for
example, with the PACS instrument \citep{Poglitsch-et-al-2010}
of the Herschel Space Observatory \citep{Pilbratt-et-al-2010}?
The sensitivity of the PACS instrument at $70\mum$ is $4.7\mJy$ in $1$ hour integration time
at a $5\sigma$ 
uncertainty level\footnote{\textit{http://herschel.esac.esa.int/Docs/PACS/html/ch03s05.html\#sec-photo-sensitivity} 
(Last accessed on 24 November 2011)}.
This is about a factor of $10$
above the calculated SED flux of $0.4\mJy$ at $70\mum$.
This factor would further increase when taking into account additional background noise and 
photospheric flux uncertainties. We conclude that
the detection of an exact analog of the EKB
with the present-day instruments is impossible.
An apparent contradiction to \cite{Vitense-et-al-2010}, who concluded that 
Herschel/PACS should be able to detect an $\approx 2M_\text{EKB}$ analog,
traces back mainly to a different extrapolation method from parent bodies to smallest grains.

\section{Conclusions}\label{sec:conclusion}

The purpose of this paper was to develop a self-consistent model of the
EKB debris disk.
To accomplish this task we used the debiased population of EKBOs
as described in \cite{Vitense-et-al-2010}.
Treating this population as dust parent bodies, we generated their dust disk
with our collisional code.
We draw the following conclusions:

\begin{enumerate}
 \item
We have shown that sub-kilometer-sized EKBOs largely determine the amount and distribution
of dust in the outer Solar System.
However, these are far too small to be directly detected at present in TNO 
surveys, and their properties cannot be accessed by collisional modeling,
because they are not in a collisional equilibrium.
Therefore, an extrapolation from observable TNOs towards smaller sizes is necessary.

\item
A straightforward extrapolation for the yet unknown objects ($s\lesssim10\km$) with a
classical Dohnanyi law can be ruled out. In that case, the amount of dust would be so large
that its thermal emission would have been detected by the COBE spacecraft. Therefore, the
distribution of these objects should be flatter. In other words, a
break in the size distribution at several tens of kilometers {\em has} to be present.

\item
Different extrapolation methods which are consistent with the measurements reveal the EKB either
as a transport-dominated debris disk
or to be intermediate between the collision-dominated and transport-dominated regimes.
Depending on the extrapolation method, we found the present-day EKB
to be in collisional equilibrium for objects $s<(10\dots 60)\m$.

\item
Using the results of \cite{Mustill-Wyatt-2011},
we estimated the effect of resonance trapping of planets.
The capturing rate of the dust grains that
are either detectable with in-situ measurements by 
spacecraft or contribute to measurable thermal emission
turned out to be $<10\%$ in most cases
and not to exceed $<20\%$ even for the largest grains considered.
Accordingly, resonance trapping should have a negligible effect
on dust impact rates and dust thermal emission, given the typical
accuracy of the dust measurements.

\item
Gravitational scattering of dust grains by planets
was investigated numerically.
Scattering can modify the particle flux in the Saturn-Uranus region
($8\AU<r<15\AU$) by about a factor of two and has little effect
on thermal emission of dust.

\item
Likewise, sublimation can reduce
the particle flux by approximately a factor of two and
does not affect the thermal emission fluxes perceptibly.

\item 
We calibrated our model with the in-situ measurements
of the New Horizons dust counter \citep{Poppe-et-al-2010,Han-et-al-2011}
by fitting our results to the data points
and can reproduce the nearly constant particle flux of $3\times 10^{-4}\persmpers$.
The corresponding production rate of dust inside the EKB amounts
to $2\times 10^6\gpers$, consistent with previous estimates
\citep[e.g.][]{Yamamoto-Mukai-1998,Landgraf-et-al-2002,Han-et-al-2011}.
In a steady-state collisional cascade (which we assume), the
``dust production rate'' is the same as the ``dust loss rate''.
Thus the result means that
$2$ tons of dust per second leave the system 
by inward transport and through ejection as blowout grains.

\item
The spectral energy distribution of an EKB analog, seen from a distance of $10\pc$,
would peak at $40-50\mum$ with a maximum flux of $0.5\mJy$.
This is consistent with the upper limit, placed by non-detection of thermal emission
from the EKB dust as it would be viewed from outside at $70\mum$ by the COBE spacecraft.
The fractional luminosity of the EKB was calculated to be $f_d=1.2\times 10^{-7}$.
The in-plane optical depth for $r>10\AU$ is set by our model to $2\times 10^{-6}$.
Although the Herschel/PACS instrument successfully detects debris disks
at similar fractional luminosity levels as the EKB
\citep{Eiroa-et-al-2010,Eiroa-et-al-2011},
all these are larger and therefore colder.
Their thermal emission peaks at wavelengths longward of 
$100\mum$, where 
the stellar photosphere is dimmer.
The detection of an exact EKB analog
even with PACS would not be possible.

\end{enumerate}

\section{Discussion}\label{sec:discussion}

Like every model, ours rests on many assumptions and is not free of
uncertainties. Here we discuss some issues.

\begin{enumerate}
\item Material composition.
Compared to \cite{Vitense-et-al-2010}, who applied geometric
optics in calculating the radiation pressure,
we now consistently used a more realistic material composition in 
both collisional and thermal emission calculations.
Nevertheless we assumed many parameters of solids~--- bulk density, shape, porosity,
tensile strength, and others~--- to be the same across the entire size range,
from Pluto-sized TNOs down to dust.
This assumption is obviously unrealistic.

\item The role of sub-kilometer-sized EKBOs.
Since little is known about EKBOs smaller than a few
tens of kilometers,
but these largely control the amount and distribution
of dust in the outer Solar System,
an extrapolation from observable TNOs towards smaller sizes is necessary.
The question is what kind of extrapolation is reasonable.
If the parent bodies inherit their orbital elements
to their children and grand-children, then the EKB should comprise a huge amount
of meter- and sub-kilometer-sized objects in highly eccentric orbits, stemming from
scattered EKBOs.
This would make collisions more disruptive and alter the size distribution of dust.
The resulting size distribution would be dominated by the smallest dust grains,
just above the radiation pressure blowout limit.
If, in contrast, the meter- and sub-kilometer-sized objects have moderate eccentricities,
the peak of the cross-section in the size distribution would be broader and shifted to larger 
grains. To distinguish between these possibilities, one needs
more information about the amount and distribution of sub-kilometer objects.
Accurate measurements of sizes and orbital elements
of dust grains in the outer Solar System in the future would also help.

\item Break in the size distribution of EKBOs.
Our model shows that a break in the size distribution at tens of kilimeters, as reported is 
the recent literature, is necessary. Otherwise the EKB dust disk would be too
dusty, violating the available observational constraints.
If such a break is present in other debris disks as well,
then the total mass of parent bodies should be higher that that usually inferred
in the debris disks studies.

\item Planetary scattering and sublimation.
If planetary scattering and/or sublimation is more efficient than assumed in our model,
the amount of dust grains that reach the Saturn-Uranus region of the Solar System 
would be smaller.
To stay consistent with the in-situ measurements, one would have to compensate
higher scattering rates and/or more efficient sublimation
by higher dust production rates in the EKB.
However, this would lead to an higher thermal emission flux,
which would contradict the non-detection of thermal emission by COBE.

\end{enumerate}

More and deeper TNO surveys,
including good measurements of their orbital elements (in particular the eccentricity), 
would help to improve the
extrapolation method and therefore our model, resulting in tighter constraints
on the dust distribution and more accurate predictions for the upcoming dust flux 
measurements. 
Better constraints on the
population of sub-kilometer objects, which could be expected, for instance, from
the stellar occultation method 
\citep[e.g.,][]{Liu-et-al-2008,Schlichting-et-al-2009,Bianco-et-al-2010},
would also be of great help.
Of course, the most promising way to improve the model of the EKB dust is to use 
direct observations of dust. In particular, a size distribution of the impacted grains
on the New Horizons dust counter would be very helpful as well as new thermal
emission contraints that could be expected from the Planck mission \citep{Ade-et-al-2011}.

\begin{acknowledgements}
We would like to thank Mih\'aly Hor\'anyi for providing us with new data of the New Horizons 
dust counter and helpful discussions of several aspects of this work.
Useful comments of the anonymous referee are very much appreciated.
This research was supported by the
\emph{Deut\-sche For\-schungs\-ge\-mein\-schaft} (DFG), projects number Kr~2164/9-1
and Lo~1715/1-1.
\end{acknowledgements}


\newcommand{\AAp}      {Astron. Astrophys.}
\newcommand{\AApSS}    {AApSS}
\newcommand{\AApT}     {Astron. Astrophys. Trans.}
\newcommand{\AdvSR}    {Adv. Space Res.}
\newcommand{\AJ}       {Astron. J.}
\newcommand{\AN}       {Astron. Nachr.}
\newcommand{\AO}       {App. Optics}
\newcommand{\ApJ}      {Astrophys. J.}
\newcommand{\ApJS}     {Astrophys. J. Suppl.}
\newcommand{\ApJL}     {Astrophys. J. Letters}
\newcommand{\ApSS}     {Astrophys. Space Sci.}
\newcommand{\ARAA}     {Ann. Rev. Astron. Astrophys.}
\newcommand{\ARevEPS}  {Ann. Rev. Earth Planet. Sci.}
\newcommand{\BAAS}     {BAAS}
\newcommand{\CelMech}  {Celest. Mech. Dynam. Astron.}
\newcommand{\EMP}      {Earth, Moon and Planets}
\newcommand{\EPS}      {Earth, Planets and Space}
\newcommand{\GRL}      {Geophys. Res. Lett.}
\newcommand{\JGR}      {J. Geophys. Res.}
\newcommand{\JQSRT}    {J. Quantitative Spectroscopy and Radiative Transfer}
\newcommand{\MNRAS}    {MNRAS}
\newcommand{\PASJ}     {PASJ}
\newcommand{\PASP}     {PASP}
\newcommand{\PSS}      {Planet. Space Sci.}
\newcommand{\SolPhys}  {Sol. Phys.}
\newcommand{\SolSysRes}{Sol. Sys. Res.}
\newcommand{\SSR}      {Space Sci. Rev.}

\end{document}